\documentclass[twocolumn,preprintnumbers,amsmath,amssymb,prb,superscriptaddress]{revtex4}

\usepackage{graphicx}
\usepackage{ulem}
\usepackage{color}
\usepackage{float}
\usepackage[colorlinks=true,plainpages=false,linkcolor=blue,urlcolor=blue,citecolor=blue,pdfpagemode=UseNone,pdfstartview=FitBH]{hyperref}

\def\CRO{Ca$_2$RuO$_4$}

\begin{document}

\title{Observation of spin-orbit excitations and Hund's multiplets in \CRO}
\author{H.~Gretarsson}
\affiliation{Max-Planck-Institut f\"{u}r Festk\"{o}rperforschung, Heisenbergstr. 1, D-70569 Stuttgart, Germany}
\affiliation{Deutsches Elektronen-Synchrotron DESY, Notkestr. 85, D-22607 Hamburg, Germany}
\author{H.~Suzuki}
\affiliation{Max-Planck-Institut f\"{u}r Festk\"{o}rperforschung, Heisenbergstr. 1, D-70569 Stuttgart, Germany}
\author{Hoon Kim}
\affiliation{Max-Planck-Institut f\"{u}r Festk\"{o}rperforschung, Heisenbergstr. 1, D-70569 Stuttgart, Germany}
\affiliation{Department of Physics, Pohang University of Science and Technology, Pohang 790-784, Republic of Korea}
\affiliation{Center for Artificial Low Dimensional Electronic Systems, Institute for Basic Science (IBS), 77 Cheongam-Ro, Pohang 790-784, Republic of Korea}
\author{K.~Ueda}
\author{M.~Krautloher}
\affiliation{Max-Planck-Institut f\"{u}r Festk\"{o}rperforschung, Heisenbergstr. 1, D-70569 Stuttgart, Germany}
\author{B. J.~ Kim}
\affiliation{Max-Planck-Institut f\"{u}r Festk\"{o}rperforschung, Heisenbergstr. 1, D-70569 Stuttgart, Germany}
\affiliation{Department of Physics, Pohang University of Science and Technology, Pohang 790-784, Republic of Korea}
\affiliation{Center for Artificial Low Dimensional Electronic Systems, Institute for Basic Science (IBS), 77 Cheongam-Ro, Pohang 790-784, Republic of Korea}
\author{H.~Yava\c{s}}
\affiliation{Deutsches Elektronen-Synchrotron DESY, Notkestr. 85, D-22607 Hamburg, Germany}
\author{G. Khaliullin}
\author{B.~ Keimer}
\affiliation{Max-Planck-Institut f\"{u}r Festk\"{o}rperforschung, Heisenbergstr. 1, D-70569 Stuttgart, Germany}

\date{\today}

\begin{abstract}
We use Ru $L_3$-edge (2838.5 eV) resonant inelastic x-ray scattering (RIXS) to quantify the electronic structure of \CRO, a layered $4d$-electron compound that exhibits a correlation-driven metal-insulator transition and unconventional antiferromagnetism. We observe a series of Ru intra-ionic transitions whose energies and intensities are well described by model calculations. In particular, we find a $\rm{J}=0\rightarrow 2$ spin-orbit excitation at 320 meV, as well as Hund's-rule driven $\rm{S}=1\rightarrow 0$ spin-state transitions at 750 and 1000 meV. The energy of these three features uniquely determines the spin-orbit coupling, tetragonal crystal-field energy, and Hund's rule interaction. The parameters inferred from the RIXS spectra are in excellent agreement with the picture of excitonic magnetism that has been devised to explain the collective modes of the antiferromagnetic state. $L_3$-edge RIXS of Ru compounds and other $4d$-electron materials thus enables direct measurements of interactions parameters that are essential for realistic model calculations.
\end{abstract}

\maketitle

\section{INTRODUCTION}
\noindent
The influence of spin-orbit coupling (SOC) on the phase behavior of compounds with orbitally degenerate \mbox{$d$-electrons} has been a subject of intense recent interest \cite{Wit14,Rau16,Win17,Ber19}. Prominent examples include highly frustrated (``Kitaev'') exchange interactions and spin-liquid correlations in Mott insulators with strong SOC \cite{Jackeli,Cha10,Chun,Kitagawa,Tak19}, as well as profound SOC-induced modifications of the band topology and superconducting pairing interaction in $d$-electron metals \cite{Kha04,Shi09,Haverkort,Veenstra,Borisenko,Christensen}. Materials with $4d$ valence electrons are a particularly versatile platform for the exploration of SOC-driven phenomena. Next to widely studied model compounds such as the Kitaev spin-liquid candidate RuCl$_3$ \cite{Plumb,Ban17} and the unconventional superconductor Sr$_2$RuO$_4$ \cite{Sr214,Maeno}, an emerging research frontier addresses collective phenomena in $4d$-electron materials exhibiting correlation-driven metal-insulator transitions \cite{Mizokawa01,Lee,Gin13,Sof17,Higgs_Ca214,Braden17}.
Realistic modelling of these phenomena is difficult, because the SOC of $4d$-electrons is comparable in magnitude to other local interactions, including the Hund's rule and ligand-field interactions. Accurate measurements of the strength of these interactions are essential for realistic model calculations of the physical properties of $4d$-electron systems.

We have built a spectrometer for resonant inelastic x-ray scattering (RIXS) that allows direct measurements of the hierarchy of low-energy electronic interactions in $4d$-metal compounds \cite{Suzuki}. We present RIXS results on \CRO, an isovalent analogue of Sr$_2$RuO$_4$ that is based on Ru$^{4+}$ ions (electron configuration $4d^4$) in RuO$_2$ square planes.  \CRO\ has recently attracted much attention due to its Mott insulator-to-metal transition that can be driven by temperature \cite{Nakatsuji}, hydrostatic pressure \cite{Steffens1}, epitaxial strain \cite{Chris_Films}, chemical substitution \cite{Steffens2}, and electrical current \cite{Nakamura,Okazaki,Sow}.  Experiments in the insulating state revealed antiferromagnetic order with an unconventional excitation spectrum composed of a soft longitudinal (``Higgs'') mode and transverse magnons with a large gap \cite{Sof17,Higgs_Ca214,Braden17}. These data can be understood in terms of a model based on competition between the intra-atomic SOC ($\xi$) of the Ru $d$-electrons and the inter-atomic exchange interaction \cite{Gin13}. While the former imposes a non-magnetic $|\rm{J}$=0$\rangle$ ground state (where J is the quantum number for the total angular momentum), the latter promotes the condensation of $|\rm{J}$=1$\rangle$ excitons into the antiferromagnetically ordered state via a mechanism that has been termed ``excitonic magnetism'' \cite{Gin13}. The tetragonal crystal field of strength $\Delta$ acting on the Ru $4d$-electrons splits the degeneracy of the $|\rm{J}$=1$\rangle$ manifold and extends the stability range of antiferromagnetism. Even in its insulating state, the phase behavior of  \CRO\ is thus controlled by a delicate balance between different interactions that have to be determined experimentally to arrive at a microscopic understanding of the magnetic ground state and excitations. To understand the insulator-metal transition and the multiple instabilities in the metallic state, the Hund's rule interaction, $\rm{J_H}$, is also of crucial importance. 

Using RIXS at the dipole-active Ru $L_3$-edge (2838.5 eV), we have uncovered a series of sharp electronic excitations in  \CRO\  from which we were able to accurately extract the parameters $\xi$, $\Delta$, and $\rm{J_H}$, in analogy to recent  Ir $L_3$-edge RIXS experiments on iridates with $5d$ valence electrons \cite{Marco_PRL_2014,YJK2017}.  In particular, we find a strong SOC-driven $\rm{J}=0\rightarrow 2$  excitation at 320 meV and directly observe Hund's-rule driven $\rm{S}=1\rightarrow 0$ spin-state transitions, split by the tetragonal crystal field, at 750 and 1000 meV. Magnetic excitations are observed at $\sim 50$ meV, consistent with neutron and Raman scattering results \cite{Sof17,Higgs_Ca214}. At higher energies (2-4 eV), multiplets corresponding to excitations from the $t_{2g}$ ground-state manifold of the Ru ions into the $e_g$ crystal-field levels are seen, so that the cubic component of the crystal-field energy, $10Dq$, can also be extracted from the RIXS spectra. The set of microscopic parameters obtained in this way specifies the low-energy Hamiltonian and places \CRO\ into the regime of excitonic magnetism \cite{Sof17,Higgs_Ca214,Gin13}. The results demonstrate the power of RIXS in elucidating the electronic structure of ruthenates and other $4d$-metal compounds. 

\section{EXPERIMENTAL DETAILS}
\begin{figure}[htb]
\includegraphics[trim=0cm 0cm 0cm 0cm, width=0.95\columnwidth]{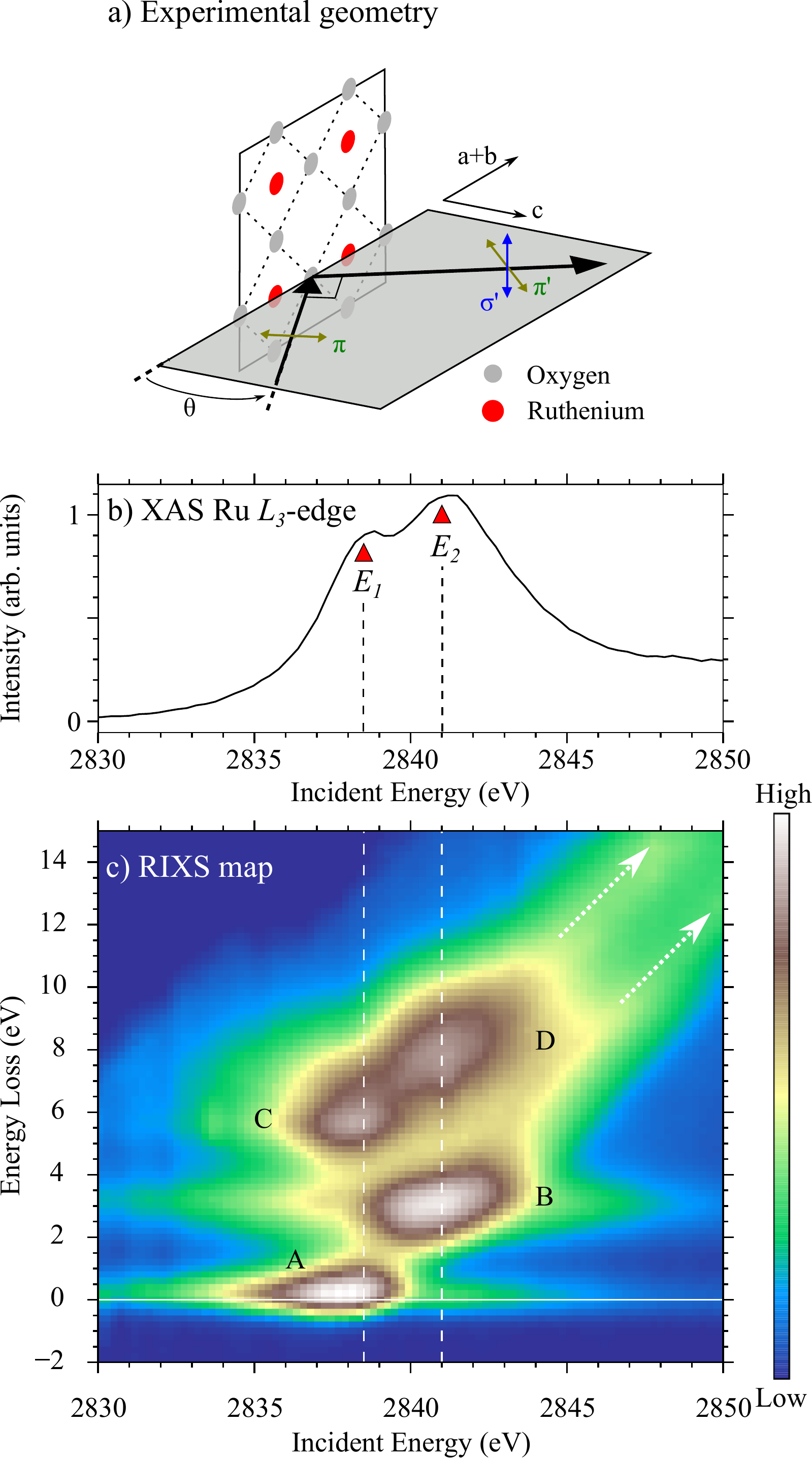}
\caption{\label{Fig_1} (a) Geometry of the RIXS experiment. The incoming and outgoing photon beams subtend a fixed angle of 90$^\circ$. By varying the angle $\theta$ between the incoming beam and the  RuO$_2$ planes of \CRO, the incoming photon polarization ($\pi$) can be changed from $E//c$ ($\theta = 0^\circ$) to $E//ab$ ($\theta= 90^\circ$).  (b) X-ray absorption spectrum (XAS) collected at the Ru $L_3$-edge of \CRO. The red triangles ($E_1$ and $E_2$) represent excitations into the empty Ru $4d$ $t_{2g}$ and $e_g$ orbitals, respectively. (c) Color map of the incident-energy dependence of the RIXS spectrum  across the Ru $L_3$-edge. The vertical white dashed lines show the resonance energies ($E_1$ and $E_2$) of features $A/C$ and $B/D$. All data were collected at room temperature.  }
\end{figure}

The RIXS experiments were carried out at beamline P01 at the PETRA-III synchrotron at DESY, using the recently built IRIXS (Intermediate x-ray energy RIXS) spectrometer \cite{Suzuki}. A cryogenically cooled Si(111) two-bounce monochromator and a secondary Si(111) channel-cut monochromator (asymmetrically cut) were used to give an incoming bandwidth of $\sim 130$ meV at 2.840 keV. A spherical (1 m radius) diced SiO$_2$ (10$\bar{2}$) analyzer (for details on fabrication see Ref. \onlinecite{Yavas_2015})  was used to obtain an overall energy resolution of $\Delta \rm{E} \sim 160$~meV (for details see Appendix \ref{app:Energy}). A single crystal of \CRO\ was grown by the floating zone method  \cite{Nakatsuji_Growth_2001}. The lattice parameters of $a = 5.4$\AA, $b = 5.5$\AA, and $c = 11.9$\AA\ were determined by x-ray powder diffraction, in good agreement with the parameters reported in the literature \cite{Braden_PRB_1998}. Due to twin domains we do not distinguish between $a$- and $b$-axes.  The magnetic ordering temperature $T_{\rm{N}} = 110$ K was determined by magnetometry.  The RIXS experiment was carried out using the geometry displayed in Fig. \ref{Fig_1} (a), and the temperature was kept at 12 K unless stated otherwise. The crystal was mounted in the $[H,H,L]$ scattering plane (orthorhombic unit cell). The outgoing photons were detected at a fixed angle of $90^\circ$ with respect to the incoming photons. To determine the energy of the elastic line, we measured scattering from a carbon tape placed adjacent to the sample.

\begin{figure*}[htb]
\includegraphics[trim=0cm 0cm 0cm 0cm, width=2\columnwidth]{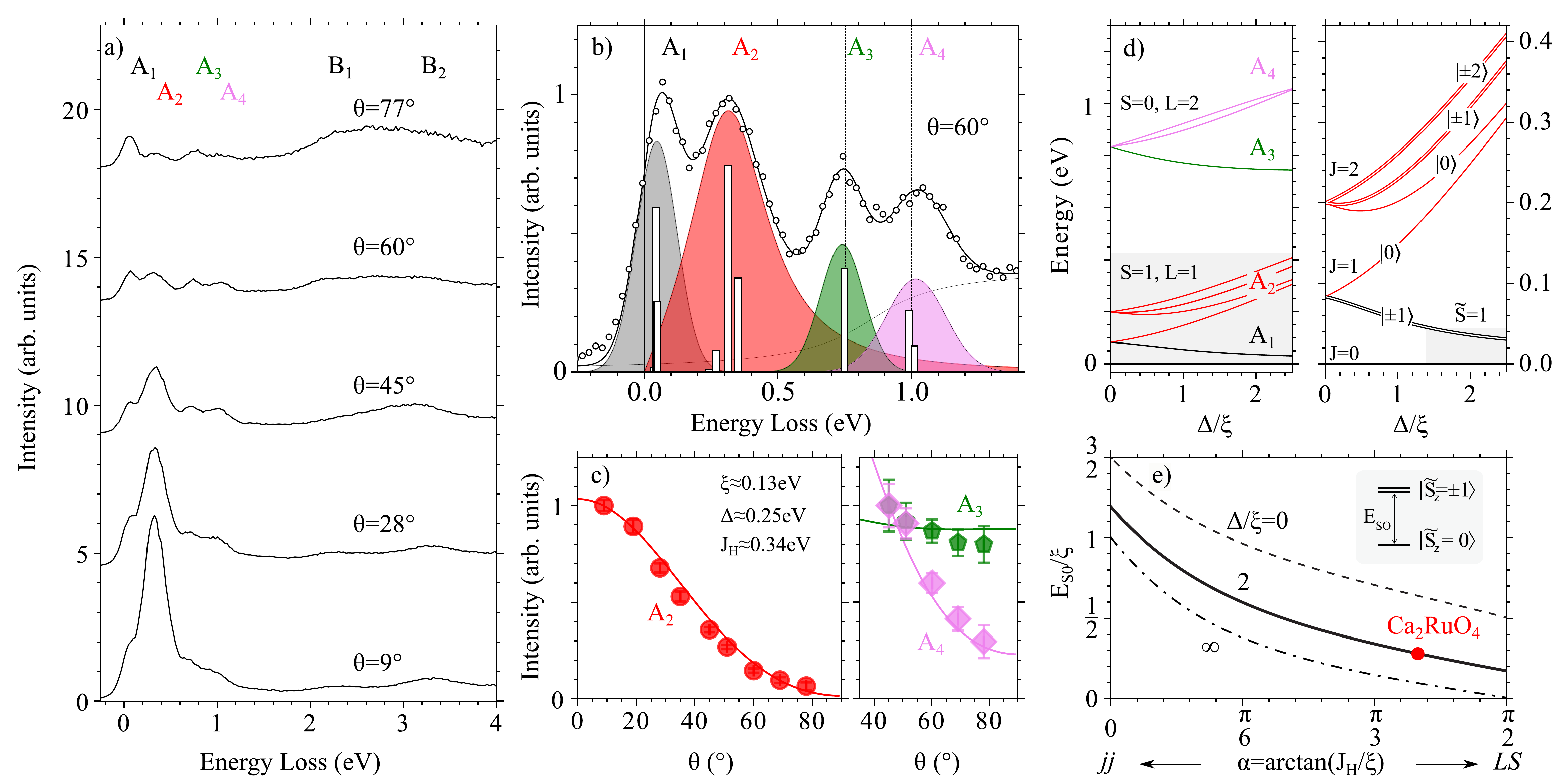}
\caption{\label{Fig_2} (a) RIXS spectrum of \CRO\ collected for incident energy $E_1 = 2838.5$ eV. The sample was kept at a temperature of 12 K while varying the incident angle $\theta$. Each spectrum was shifted vertically for clarity. (b) Fitted RIXS spectrum taken at $\theta=60^\circ$.  The vertical white bars are a result of multiplet calculations (see text). (c)  Intensity of the A$_2$ (A$_3$ and A$_4$) feature as a function of $\theta$. The data were normalized to the value at $\theta=9^\circ$ (45$^\circ$); solid curves represent theory results. (d) The $t_{2g}^4$ multiplet energy levels as a function of $\Delta/\xi$. The SOC was fixed at $\xi=0.13$ eV. Hund's coupling separates the multiplet into levels with different spin S and orbital angular momentum L. The shaded region below 0.4 eV is magnified on the right hand side for the detailed structure of the $\rm{S}=1$ states. For large values of $\Delta$, the lowest three levels (J=0 ground state, and $\rm{J_z}=\pm$1 doublet at E$_{\rm{so}}$) form an effective $\tilde{S}$=1 system hosting excitonic magnetic order \cite{Gin13,Higgs_Ca214}. e) Spin-orbit excitation energy E$_{\rm{so}}$ (see inset) as a function of $\rm{J_H}/\xi$ at different values of the tetragonal crystal field $\Delta/\xi$. The limiting case of $\rm{J_H}/\xi =0$ ($\infty$) corresponds to the so-called $jj$-coupling ($LS$-coupling) scheme. The location of \CRO\ with $\rm{J_H}/\xi\simeq 2.6$ and $\Delta/\xi \simeq 2$ is indicated.}
\end{figure*}

\section{EXPERIMENTAL RESULTS}
\subsection{Incident energy dependence}

Figure~\ref{Fig_1}(b) shows the Ru $L_3$-edge x-ray absorption spectrum of \CRO. The data were collected at room temperature in the total fluorescence yield mode. The sample normal ($c$-axis) subtended an angle of $\theta=30^\circ$ with the incoming photon polarization. Two features can be observed at incident energies of $E_1 = 2838.5$ eV and $E_2 = 2841$ eV, corresponding to the $2p_{3/2} \rightarrow 4d$ $ t_{2g}$ and $2p_{3/2} \rightarrow 4d$ $e_{g}$ transition, respectively. The splitting between these two features (2.5 eV) is in good agreement with results on other Ru $d^4$ systems \cite{Tjeng_PRL_2006}. 

In Fig.~\ref{Fig_1}(c) we plot the incident energy ($E_i$) dependence of a low-resolution RIXS spectrum ($\Delta \rm{E} \sim 900$~meV) across the Ru $L_3$-edge.  For $E_i = E_1$, features $A$ and $C$ display resonances, while $B$ and $D$ resonate at $E_i = E_2$  (dashed vertical white lines). This observation shows that $A$ and $C$ ($B$ and $D$) originate from transitions into the same unoccupied $t_{2g}$ ($e_{g}$) manifold, but differ in their final state. We can thus identify $A$ and $B$ as ``$dd$-excitations'' originating from intra-$t_{2g}$ and $t_{2g} \rightarrow e_{g}$ excitations, respectively, while $C$ and $D$ most likely originate from charge-transfer excitations. Above the Ru $L_3$ edge ($E_i>2845$ eV), the L$\beta_{2,15}$ emission line of \CRO\ is seen (dashed white arrows). 

\subsection{Fine structure and polarization dependence}

The data in Fig.~\ref{Fig_1}(c) imply that excitations within the $t_{2g}$ multiplets [feature $A$ in Fig.~\ref{Fig_1}(c)] are resonantly enhanced at $E_i=E_1$ and appear only below 1 eV. Armed with this result, we can now study the fine structure of feature $A$. In Fig.~\ref{Fig_2}(a), high-resolution RIXS spectra ($\Delta \rm{E} \sim 160$ meV) of \CRO\ are plotted for multiple $\theta$ values. Each spectrum was normalized to the intensity in the featureless spectral region between 1.3 and 1.5 eV \cite{selfabs,Troger_PRB_1992}. Inspection of the spectrum collected at $\theta=60^\circ$ reveals that feature $A$ consists of four components ($A_1 - A_4$): a quasi-elastic line followed by three peaks between 0 and 1 eV. We note that the lack of strong elastic scattering is not unexpected and is an indicator of good crystal quality \cite{Jungho_PRL_2012}. At higher energy losses an electronic continuum appears \cite{Tokura_PRL_2003}, followed by the $e_g$ multiplets $B_1$ and $B_2$ at $\sim$2.3 and 3.3 eV, respectively. By lowering $\theta$ the spectrum changes dramatically. The intensity of feature $A_2$, which is comparable to the other features at higher $\theta$, becomes dominant. It is also clear that $A_2$ exhibits an asymmetric lineshape that likely originates from more than one excitation whose splitting is below our resolution.

To extract the energies and intensities of the spectral features from the high-resolution RIXS data in Fig.~\ref{Fig_2}(a), we fitted each spectrum using a superposition of four profiles. A set of three Gaussian functions were used to represent $A_1$, $A_3$ and $A_4$, while an antisymmetric Lorentzian was used to model the asymmetric $A_2$ feature. An example of the fit can be seen in Fig.~\ref{Fig_2}(b). The solid black line is the result of the complete fit, and filled features represent the individual contributions. The fit yields energies of 50 meV for $A_1$, 320 meV for $A_2$ and 750 (1000)~meV for $A_3$ ($A_4$) (see Appendix \ref{app:Energy}). We note that within the energy resolution of our instrument the peaks do not disperse when varying the momentum transfer via the incident angle $\theta$. 

We now address the photon polarization dependence of the RIXS intensity, which is also modulated by $\theta$ and provides additional clues to the origin of the different features. When increasing $\theta$, the polarization of the incoming photon moves from the sample $c$-axis into the $ab$-plane. Fig.~\ref{Fig_2}(c) shows the intensity of feature $A_2$ (normalized to the value at $\theta=9^\circ$) as a function of $\theta$.  The plot thus clearly demonstrates a strong polarization dependence of the $A_2$ intensity. On the right hand side we show a similar plot for $A_3$ and $A_4$, where the intensity was normalized to its value at $\theta=45^\circ$ \cite{note2}. Different from the $A_2$ feature, $A_3$ largely retains its intensity while $A_4$ shows suppression with increasing $\theta$. 

\section{THEORETICAL INTERPRETATION}

To gain insight into the origin of the multiple features seen in our RIXS spectra, we have carried out ionic model calculations that quantify the energy levels of Ru $d^4$ multiplets and corresponding RIXS intensities. The Hamiltonian we use is standard and includes intra-ionic Hund's coupling $\rm{J_H}$, spin-orbit coupling $\xi$, tetragonal $\Delta$ and cubic $10Dq$ crystal field splittings; see Appendix \ref{app:Hamiltonian} for its explicit form. 

The ionic model has a rich multiplet structure but is local in space, so it can be easily diagonalized numerically for arbitrary parameter values. In Fig.~\ref{Fig_2}(d) we plot the calculated energies of the $t_{2g}^4$ multiplets as a function of $\Delta$/$\xi$. Hund's rule selects $|\rm{S}$=1,$\rm{L}$=1$\rangle$ as the lowest level of the $t_{2g}^4$ manifold (black and red levels). At higher energies the system accommodates low-spin states $|\rm{S}$=0,$\rm{L}$=2$\rangle$ split by the tetragonal crystal field $\Delta$ (green and violet levels). The SOC splits the $\rm{S}=1$ manifold into $|\rm{J}$=0$\rangle$, $|\rm{J}$=1$\rangle$, and $|\rm{J}$=2$\rangle$ states, as detailed in the right panel of Fig.~\ref{Fig_2}(d). The tetragonal compression then brings the $|\rm{J}$=1,$\rm{J_z}$=$\pm$1$\rangle$ doublet close to the ground state singlet $|\rm{J}$=0$\rangle$, forming a three-level structure that can be described by an effective $\rm{\tilde{S}} = 1$ low-energy model \cite{Sof17, Higgs_Ca214}.  The energy of the remaining $|\rm{J}$=1, $\rm{J_z}$=0$\rangle$ state is raised by the compression, close to the $|\rm{J}$=2$\rangle$ states. The energy of the $|\rm{J}$=2$\rangle$ states results from combined action of $\xi$ and $\Delta$ and gradually increases with $\Delta/\xi$. 

Based on the energy diagram in Fig.~\ref{Fig_2}(d), we can assign the $A_1$ peak to magnetic transitions within the low-energy singlet-doublet sector (black lines), $A_2$ to spin-orbit $\rm{J}=0\rightarrow 2$ excitations (red lines), while $A_3$ and $A_4$ originate from $\rm{J_H}$-driven spin-state transitions split by the tetragonal field $\Delta$. The peak positions depend sensitively on $\Delta$, $\xi$, and $\rm{J_H}$, and an excellent fit is obtained for $\Delta=0.25$~eV, $\xi=0.13$~eV, and $\rm{J_H}=0.34$~eV. Note that the number of observable spectral features $A_2$, $A_3$, and $A_4$ uniquely determines all three parameters.

As a consistency check of the above assignment, we have also calculated the RIXS intensities of the transitions in Fig.~\ref{Fig_2}(a), based on the scattering geometry in Fig.~\ref{Fig_1}(a). We used numerically obtained multiplet wavefunctions and adopted the fast collision approximation \cite{fast_rixs} for the RIXS operator (see Appendix \ref{app:Intensity} for details). The results obtained for $\theta=60^\circ$ can be seen in Fig.~\ref{Fig_2}(b) as white vertical bars, and the polarization dependence of these transitions is plotted as solid lines in Fig.~\ref{Fig_2}(c). Overall the calculations agree very well with the experimental observations. In particular, the $\rm{J}=0\rightarrow 2$ transitions $A_2$ are strongly $\theta$-dependent, as observed, and the bifurcating behavior of the spin-state transitions is reproduced -- the lower peak $A_3$ largely maintains its intensity with increasing $\theta$, while the upper peak $A_4$ diminishes, see Fig.~\ref{Fig_2}(c). 

To describe the dispersive magnons and amplitude (Higgs) mode that give rise to the low-energy $A_1$ feature, one has to go beyond the local model. We have adopted the effective $\rm{\tilde{S}}=$1 model of Refs.~\onlinecite{Higgs_Ca214,Sof17} that is built on the low-energy singlet-doublet sector of the $d^4$-ion, see Fig.~\ref{Fig_2}(d). The corresponding Hamiltonian can be represented as 

\begin{align}
\label{eff_S}
H_{\tilde{S}} = J \sum_{\langle ij \rangle}{ \bm{\tilde{S}}_i \cdot \bm{\tilde{S}}_j }+ E_{so}\sum_i { \tilde{S}_{zi}^2 }\;, 
\end{align}

\noindent neglecting small anisotropy terms~\cite{Higgs_Ca214,Sof17} which are not relevant here. The exchange interaction $J$ triggers a condensation of $\tilde{S}_z=\pm$1 states, driving the system into a magnetically ordered phase. The excitation spectra of the model have been derived earlier, see Ref.~\onlinecite{Higgs_Ca214}. We recovered the $A_1$ feature at $\sim$~50 meV by using the parameters $J=$ 5.8 meV and $\rm{E_{so}} =$ 27 meV \cite{Higgs_Ca214}. The intensities of the magnon and Higgs modes have been calculated using the RIXS operators for the $\rm{\tilde{S}}=$~1 model, given by Eq.~(31) of Ref.~\onlinecite{Kim17}. For the wave-vectors accessed in the current scattering geometry, these collective modes are found to have a moderate intensity, comparable to that of the local spin-orbital transitions $A_2$-$A_4$, consistent with observations. In general, however, the theoretical calculations (to be presented elsewhere \cite {Hoon19}) show that the RIXS intensity of these modes should be strongly enhanced near the magnetic Bragg points. 

The spin-orbit induced energy gap E$_{\rm{so}}$ between the $\rm{J}=0$ singlet ground state and the excited magnetic doublet $|\pm1\rangle$ in Fig.~\ref{Fig_2}(d) is a crucial parameter, which, in competition with the exchange interactions, determines magnetic ordering in Ca$_2$RuO$_4$~\cite{Gin13,Higgs_Ca214}. In general, the spin gap E$_{\rm{so}}$ depends on $\xi$, $\Delta$, and $\rm{J_H,}$ as illustrated in Fig.~\ref{Fig_2}(e). This figure shows that the spin-orbit induced magnetic gap E$_{\rm{so}}$ is always nonzero for $d^4$ ions, except in the unrealistic limit of $\Delta/\xi=\infty$ and $\rm{J_H}/\xi=\infty$ ($LS$-coupling limit). With $\rm{J_H}/\xi=2.6$ as obtained above for Ca$_2$RuO$_4$, we find that corrections to the $LS$-coupling scheme are sizeable in ruthenates, raising E$_{\rm{so}}$ by a factor of $\sim 3/2$ from the value that would follow from the $LS$-coupling approximation. This figure also suggests that iridates with $\rm{J_H}/\xi \sim 0.6$ ($\alpha \sim \pi/6$) are actually closer to the $jj$-coupling regime.

Having fully quantified our RIXS data below 1 eV, we now discuss the higher-energy spectra which show a two-peak structure, $B_1$ and $B_2$, evolving into a broader single peak at higher values of $\theta$, see Fig.~\ref{Fig_2}(a). This segment of the spectra is dominated by multiplet transitions between the $t_{2g}^4$ and $t_{2g}^3e_g$ electronic configurations, which can be readily analyzed within the ionic model (see the Hamiltonian in Appendix \ref{app:Hamiltonian}). 

\begin{figure}[htb]
\includegraphics[trim=0cm 0cm 0cm 0cm, width=0.95\columnwidth]{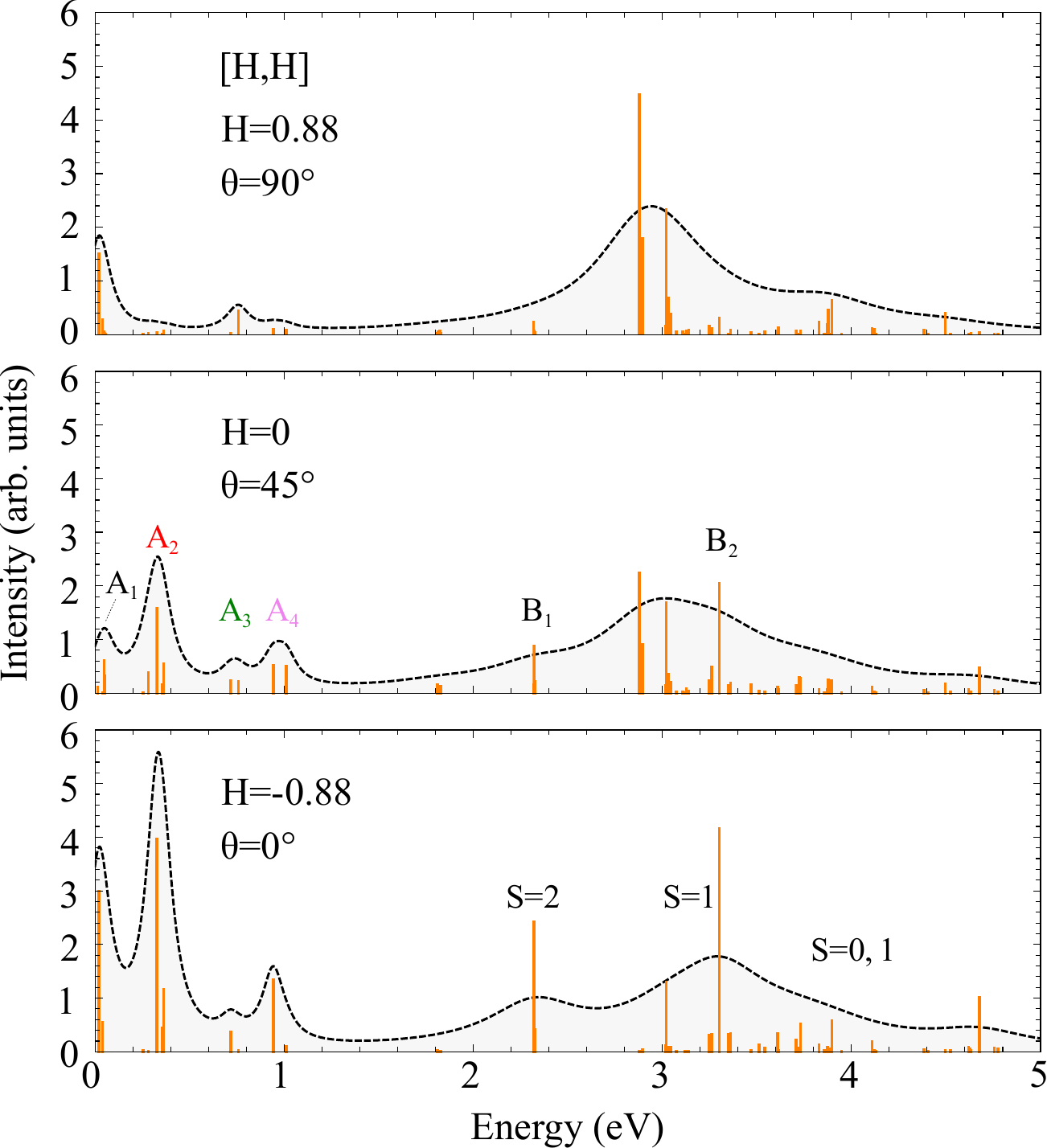}
\caption{\label{Fig_3}The calculated RIXS spectra including magnetic excitations $A_1$, spin-orbital excitations $A_2$, $A_3$, and $A_4$ within the $t_{2g}$-multiplets, and $t_{2g} \rightarrow e_{g}$ transitions $B_1$ and $B_2$. Vertical bars represent the energy and intensities of the transitions, and their Lorentzian-broadened profiles result in the dashed curves. Different widths 0.14 eV and 0.6 eV are used for the transitions within $t_{2g}$ and $e_g$ sectors, respectively. The scattering momenta $[H,H]$ and corresponding $\theta$ values are indicated.}
\end{figure}

The calculated spectra in a broad energy window including the $t_{2g} \rightarrow e_g$ transitions are shown in Fig.~\ref{Fig_3}. We used $10Dq=3.1$ eV \cite{note3} and an $e_g$ orbital splitting $\Delta_e$ = 2$\Delta$. The $t^4_{2g}\rightarrow t^3_{2g}e_g$ multiplet transitions are widely spread over the energy window of $\sim 2 - 5$~eV. The two-peak structure at $\sim$ 2.3 eV and $\sim$ 3.3 eV is clearly developed at small scattering angles $\theta$, in a qualitative agreement with the experimental data in Fig.~\ref{Fig_2}(a). We can assign the $B_1$ and $B_2$ features to the high-spin $|t_{2g}^3e_g$,~S=2$\rangle$ levels and the $|t_{2g}^3e_g$,~S=1$\rangle$ states, respectively, see the spin-state labels in Fig.~\ref{Fig_3}. The splitting between both features is 3 - 4 $\rm{J_H}$. While increasing $\theta$ suppresses the lower peak $B_1$, the higher-energy $B_2$ peak (S=1 states) transfers its spectral weight to lower energies, gently shifting its position. We expect that strong Jahn-Teller coupling of $e_g$-electrons to the lattice, and also coupling to the underlying electronic continuum above the Mott gap \cite{Tokura_PRL_2003} should substantially broaden the $t_{2g}\rightarrow e_g$ high-energy transitions. We also note that the calculated spectral weight of the $e_g$-multiplets is seemingly stronger than observed in the experiment, because the resonance profile is not explicitly accounted for in our RIXS-intensity calculations. 

\section{DISCUSSION}

We now compare our results to previous experimental and theoretical work. The Hund's rule coupling constant 0.34 eV extracted from our RIXS results is comparable to the value $\rm{J_H} \sim 0.4$ eV that was obtained by fitting angle-resolved photoemission spectroscopy data to a model based on dynamical mean-field theory \cite{Dsu17}. We note that $\rm{J_H}$ is considerably larger than that in iridates where $\rm{J_H}\sim 0.25$ eV  \cite{YJK2017}. As a consequence of covalency, the spin-orbit coupling parameter $\xi= 0.13$ eV is reduced from its free-ion value of $\xi_0= 0.16$ eV by a so-called covalency factor $\kappa\simeq$ 0.81, which is typical for Ru-ions \cite{Abragam}. The tetragonal splitting $\Delta$ is in reasonable agreement with first-principles calculations  \cite{Fang04, Gorelov10}.

In recent O $K$-edge RIXS experiments \cite{Chang_PRB_2015,Chang_PRX_2018} a low energy spin-orbit excitation was observed around 350 meV, in reasonable agreement with our results. However, the spin-state transitions at 750 and 1000 meV were not detected, most likely due to their low intensities. As a result, these transitions were instead assigned to a single broad feature at $1.3$ eV, resulting in a large $\rm{J_H}\sim 0.5$ eV  \cite{Chang_PRX_2018}. This difference is of crucial importance since the spin-state transitions are split by the tetragonal crystal field, and their observation is required to obtain absolute values of $\Delta$ and $\xi$.

With the values of $\Delta$, $\xi$, and $\rm{J_H}$ obtained in our experiment, and using the ionic model results in Fig.~\ref{Fig_2}(e), we can estimate the singlet-doublet gap $\rm{E_{so}} \simeq$ 36 meV, to be compared with a single-ion anisotropy term in the effective $\rm{\tilde{S}}=1$ model. Our estimate is somewhat larger than that deduced from neutron scattering (27 meV) and Raman (31 meV) data \cite{Higgs_Ca214,Sof17}. The difference may originate from renormalization of $\rm{E_{so}}$ by effects beyond the ionic model, and/or due to softening of spin-orbit exciton energy $\rm{E_{so}}$ by electron-phonon interactions \cite{HLiu_PRL} that are not included in our calculations. 

We observed that the transverse magnons and the Higgs mode contribute to peak $A_1$ at $\sim$~50 meV, consistent with Ref.~\onlinecite{Higgs_Ca214}. However, these modes could not be individually resolved here due to insufficient energy resolution. We note also that in the $[H,H,L]$ scattering plane used in our experiment, the Bragg peaks are not accessed and thus these modes do not show strong dispersion or intensity variations.

\section{CONCLUSIONS}
We have presented an experimental investigation of \CRO\ using a newly built Ru $L_3$-edge RIXS spectrometer, and we quantified its basic electronic parameters $\rm{J_H}$, $\xi$, $\Delta$, and $10Dq$. The parameters we obtained confirm the spin-orbit entangled nature of the low-energy states, lending strong support to the picture of excitonic magnetism in \CRO\ \cite{Gin13,Higgs_Ca214,Sof17}. More generally, our findings will encourage further RIXS studies of \CRO\ and other $4d$-metal compounds, including the emergence of new phases in doped single crystals \cite{Ca214_Doping,Chaloupka_Doping} and the magnetic dynamics in strained thin films \cite{Chris_Films}.

\acknowledgements{We would like to thank D. Ketenoglu and M. Harder for the fabrication of the SiO$_2$ analyzer,  J. Bertinshaw, H. Liu, C. Dietl, and H.C. Wille for fruitful discussions and S. Mayer and F.-U. Dill for technical help. The project was supported by the European Research Council under Advanced Grant No. 669550 (Com4Com). We acknowledge DESY (Hamburg, Germany), a member of the Helmholtz Association HGF, for the provision of experimental facilities.}

\appendix

\section{Energy Resolution and Peak positions}
\label{app:Energy}

\begin{figure}[htb]
\includegraphics[trim=0cm 0cm 0cm 0cm, width=0.95\columnwidth]{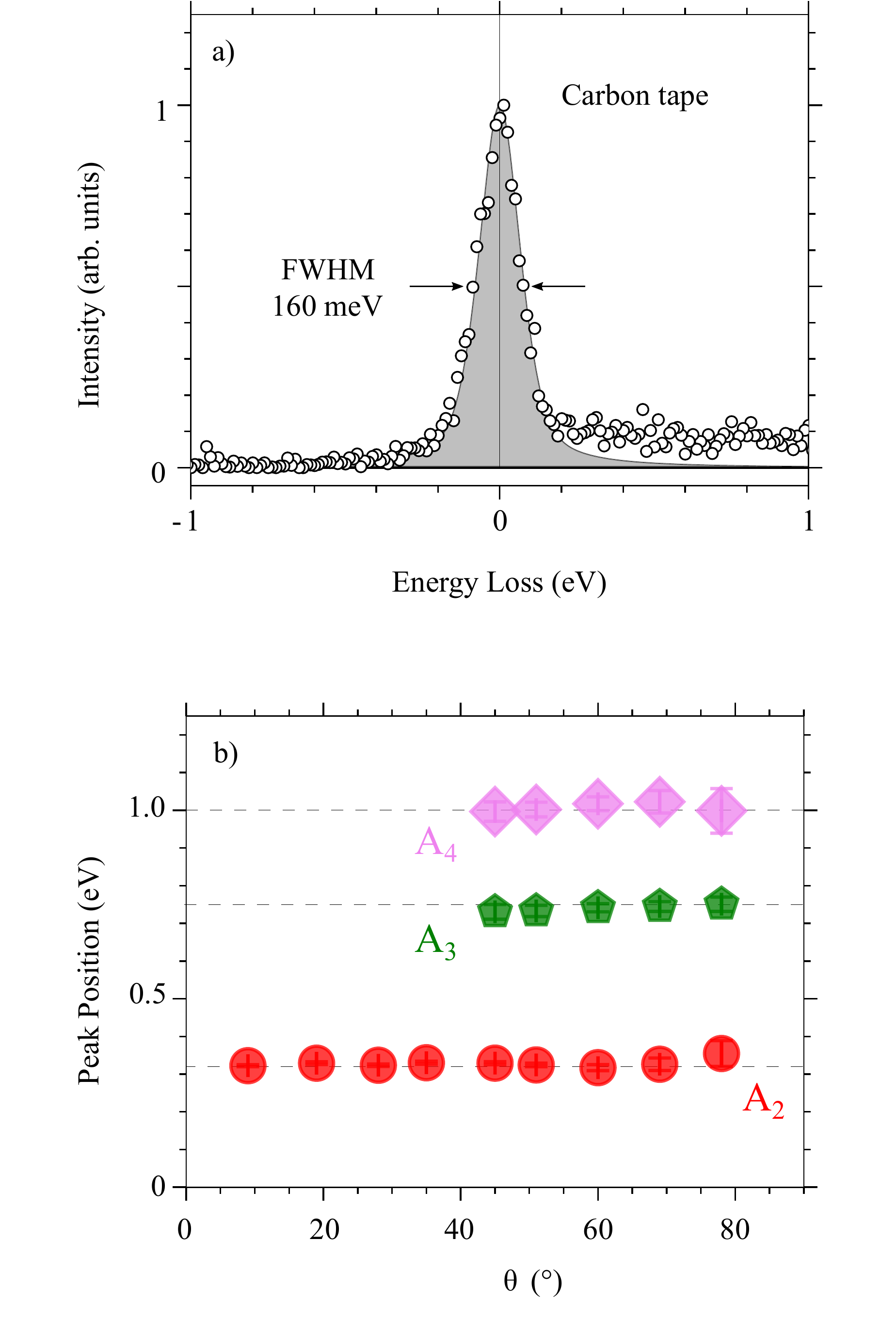}
\caption{\label{Fig_4} (a) RIXS spectrum of a carbon tape demonstrating the resolving power at 2.839 keV.  (b) Additional results for the low energy RIXS fit. Peak positions of $A_2$, $A_3$, and $A_4$ as a function of the angle $\theta$. The data points are all found within  an energy of 320$\pm 15$ meV , 750$\pm 15$ meV , and 1000$\pm 15$ meV (dashed horizontal lines).  }
\end{figure}

The total energy resolution of the IRIXS instrument can be approximated using the following formula: 

\begin{align}
\Delta\rm{E} \approx \sqrt{\Delta\rm{E_i}^2+\Delta\rm{E_a}^2+\Delta\rm{E_g}^2},
\end{align}

\noindent where $\Delta \rm{E_i}=130$ meV is the incoming x-ray bandwidth, $\Delta \rm{E_a}=60$ meV is the intrinsic resolution of the SiO$_2$(10$\bar{2}$) analyzer, and $\Delta \rm{E_g}\approx 60$ meV is the general geometric contribution, including the Johann error \cite{MMSala_2018}. This approximation gives $\Delta \rm{E} \approx 160$ meV and is in excellent agreement with our measurements of the elastic line from a carbon tape [Fig.~\ref{Fig_4}(a)]. The data in Fig.~\ref{Fig_4}(a) were fitted using a pseudo-Voigt function which consists of a linear combination of Lorentzian and Gaussian profiles with equal widths and amplitudes. 

In Fig.~\ref{Fig_4}(b) we plot the peak positions of features $A_2$, $A_3$, and $A_4$ as a function of the angle $\theta$. The two extreme data points for $A_2$ correspond to a momentum transfer $q \approx\pm(0.7,0.7)$ (orthorhombic notation). We were not able to observe any dispersion within our instrumental resolution. The dashed horizontal lines represent the peak positions (320, 750, and 1000 meV) used for comparison with the theoretical calculations. 

\section{Ionic Model Hamiltonian}
\label{app:Hamiltonian}

The energy levels and multielectron wavefunctions $|d^n\rangle$ of the Ru-ion are obtained by diagonalizing the Hamiltonian which includes the intra-ionic Coulomb interactions, spin-orbit coupling, and crystal fields. The Coulomb interaction is expressed in terms of Kanamori parameters $U$, $U'$, and $\rm{J_H}$ \cite{Sug70, Geo13} as follows:

\begin{align}
H_C &= U \sum_m{n_{m\uparrow}n_{m\downarrow}} + U'\sum_{m\neq m'}{n_{m\uparrow} n_{m' \downarrow}} \notag \\ &+ (U' - J_H) \sum_{m<m', \sigma}{n_{m\sigma}n_{m'\sigma}} \notag \\
&-J_H\sum_{m \neq m'} {d^{\dagger}_{m\uparrow}d_{m\downarrow}d^{\dagger}_{m'\downarrow}d_{m'\uparrow}}  \\&+ J_H\sum_{m \neq m'}{d^{\dagger}_{m\uparrow}d^{\dagger}_{m\downarrow}d_{m'\downarrow}d_{m'\uparrow}}\;, \notag
\end{align}

\noindent under the widely adopted approximation $U' = U-2J_H$. Here, $U$ ($U'$) correspond to intra (inter)-orbital repulsion, and the $\rm{J_H}$ terms describe the inter-orbital Hund's exchange and pair-hopping interactions. $d^{\dagger}_{m\uparrow}$ and $n_{m\uparrow}$ are electron creation and density operators, correspondingly. The spin-orbit coupling $H_{so}$, and crystal fields of cubic $H_{cub}$ and tetragonal $H_{tetra}$ symmetries are parameterized as follows:

\begin{align}
H_{so}  &=\xi \sum_{i} \vec{l_i}\cdot\vec{s_i},\\
H_{cub} &= 10Dq \cdot[\; \frac{3}{5} n_{e_g} - \frac{2}{5} n_{t_{2g}} \;],\\
H_{tetra} &= \frac{1}{3}\Delta(n_{xz}+n_{yz} - 2 n_{xy}) \notag\\ & +\frac{1}{2}\Delta_e (n_{z^2} - n_{x^2-y^2}). 
\end{align}

\noindent We note that the coupling constants in the above Hamiltonians are effective model parameters whose values are generally smaller than those for free-ions, because they are affected by $p$-$d$ covalency effects in a solid~\cite{Abragam}. 

Numerical diagonalization of the above Hamiltonians results in the spin-orbital energy levels discussed in the main text. The corresponding multielectron wavefunctions $|d^n\rangle$ (obtained as a linear combination of Slater determinants) are used to evaluate the RIXS matrix elements and intensities. 

\section{RIXS intensity calculations}
\label{app:Intensity}

To calculate the RIXS intensity at the Ru $L_3$-edge, we use the dipole moment operator $P=(P_x, P_y, P_z)$ which brings core electrons from the $2p_{3/2}$ level to the $4d$ states, and vice versa. Its $x$-component can be written as~\cite{Kim17}:

\begin{align}
& P_x=(d^\dagger_{zx}p_z+d^\dagger_{xy}p_y)+\frac{2}{\sqrt{3}}d^\dagger_{x^2}p_x ,
\end{align}

\noindent where $d$ and $p$ annihilate an electron in the respective orbitals. $P_{y/z}$ can be derived using cyclic permutation. For shorthand notation of the $d$ orbitals we use $d_{z^2}=d_{3z^2-r^2}$ and $d_{x^2/y^2}=-\frac{1}{2}d_{z^2}\pm\frac{\sqrt{3}}{2}d_{x^2-y^2}$.  Within the fast collision approximation \cite{fast_rixs}, the RIXS operators are described as subsequent $P_\alpha$ and $P_\beta$ dipolar transitions:

\begin{align}
\label{eq3}
&R = \sum_{\alpha, \beta} R_{\alpha\beta}\epsilon'_\alpha \epsilon_\beta ,\\
&R_{\alpha \beta} \propto \sum_{\{d^{n+1}p_{3/2}^3\}} \langle {d^n}' |P_\alpha^+ | d^{n+1} p_{3/2}^3 \rangle \langle d^{n+1} p_{3/2}^3 | P_\beta | d^n \rangle, \notag
\end{align}

\noindent where $\alpha$ and $\beta$ are $x,y,$ or $z$, and $\epsilon$ ($\epsilon'$) are the incoming (outgoing) light polarization vectors. The approximation assumes that the time dynamics of the intermediate states $|d^{n+1}p^5\rangle$ is faster than that of the low-energy final excitations of interest (e.g., magnons and spin-state transitions). This makes our calculation independent of the incident photon energy. In the present experimental scattering geometry, the light polarization vectors depend on the angle $\theta$ in the following manner:

\begin{align}
 \bm\epsilon^\pi_{in} = & (\frac{\sin\theta}{\sqrt{2}},\frac{\sin\theta}{\sqrt{2}},-\cos\theta),\notag\\  \bm\epsilon^{\pi'}_{out} =&(\frac{\cos\theta}{\sqrt{2}},\frac{\cos\theta}{\sqrt{2}},\sin\theta),\\ \bm\epsilon^{\sigma'}_{out} = &(\frac{-1}{\sqrt{2}}, \frac{1}{\sqrt{2}}, 0). \notag\;
 \end{align}

The numerically obtained multiplet wavefunctions $|d^n\rangle$ are substituted in Eq.~\ref{eq3} to calculate the RIXS intensities of the non-dispersive features $A_2$, $A_3$, and $A_4$, as well as $t_{2g} \rightarrow e_g$ transitions $B_1$ and $B_2$. To describe the $A_1$ peak, which originates from the dispersive magnons and the amplitude mode, one has to go beyond the local model; we evaluated its dispersion and intensity using the effective $\rm{\tilde{S}}=$1 model of Eq.~\ref{eff_S} in the main text.


\begin{thebibliography}{33}
\bibitem{note2}
The smaller angular range is due to difficulities in fitting the spectra for  $\theta < 45^\circ$ due to the strong intensity of the $A_1$ feature.
\end{thebibliography}
\end{document}